\documentclass[12pt]{article}
 \usepackage{epsfig}
 \def\be{\begin{equation}}
 \def\ee{\end{equation}}
 \def\bea{\begin{eqnarray}}
 \def\eea{\end{eqnarray}}
 \usepackage{graphicx}

 \catcode`\@=11
 \def\lsim{\mathrel{\mathpalette\@versim<}}
 \def\gsim{\mathrel{\mathpalette\@versim>}}
 \def\@versim#1#2{\vcenter{\offinterlineskip
 \ialign{$\m@th#1\hfil##\hfil$\crcr#2\crcr\sim\crcr } }}
 \catcode`\@=12

 \catcode`@=12
\begin{document}
 \thispagestyle{empty}
 \begin{flushright}
 UCRHEP-T613\\
 Apr 2021\
 \end{flushright}
 \vspace{0.6in}
 \begin{center}
 {\LARGE \bf Gauged Lepton Number, Dirac Neutrinos,\\ 
Dark Matter, and Muon $g-2$\\}
 \vspace{1.5in}
 {\bf Ernest Ma\\}
 \vspace{0.1in}
{\sl Department of Physics and Astronomy,\\ 
University of California, Riverside, California 92521, USA\\}
\end{center}
 \vspace{1.2in}

\begin{abstract}\
Lepton number is promoted to an $U(1)_L$ gauge symmetry in a simple extension 
of the standard model.  The spontaneous breaking of $U(1)_L$ by three units 
allows a conserved $Z^L_3$ lepton symmetry to remain, guaranteeing that 
neutrinos are Dirac fermions, which acquire naturally small masses from 
a previously proposed mechanism.  Dark matter appears as a singlet scalar, 
with dark symmetry $Z^D_3$ derivable from $Z^L_3$.  Muon $g-2$ may be 
explained.
\end{abstract}

\newpage
\baselineskip 24pt

\noindent \underline{\it Introduction}~:~ 
In the minimal standard model (SM) of quarks and leptons, it is well-known 
that baryon number $B$ and lepton number $L$ appear as accidental global 
symmetries.  It is also well-known that $B-L$ may be gauged with the 
addition of a singlet right-handed neutrino $\nu_R$ per family.
If $B$ and $L$ are allowed to be different for each family, the 
condition for an anomaly-free $U(1)_X$ gauge symmetry is simply~\cite{kmpz17}
\begin{equation}
\sum^3_{i=1} 3 n_i + n'_i = 0,
\end{equation}
where $n_i,n'_i$ are the $U(1)_X$ values for each quark and lepton family.  
If $n_i=1/3$ and $n'_i=-1$, then $B-L$ is obtained.  Many other choices 
have been considered, such as $L_\mu-L_\tau$~\cite{hjlv91,mrr02}, 
$B-3L_\tau$~\cite{m98}, etc.

The separate gauging of $B$ and $L$ is possible~\cite{fw10,dfw13,m21} with 
the addition of new fermions.  Since $\nu_R$ is also a new addition, its 
$U(1)_L$ assignment~\cite{m17} is not necessarily the same as $\nu_L$ which 
must be the same as the charged leptons.  This opens up the interesting 
option that neutrinos are naturally light Dirac fermions~\cite{mp17}.  To 
prevent $\nu_R$ from having a Majorana mass, $U(1)_L$ must not be broken by 
a singlet scalar with $U(1)_L$ charge double that of $\nu_R$.  In the 
conventional $B-L$ extension of the SM, this may be achieved with the 
breaking of gauge $B-L$ by three units~\cite{mpr13,ms15} with a scalar 
singlet $\chi^0$.  Here $\nu_R$ is chosen to transform as $-2$ under 
$U(1)_L$, so it works just as well.  To connect $\nu_L$ with $\nu_R$, 
a heavy Higgs doublet $(\eta^+,\eta^0)$ transforming as $-3$ under 
$U(1)_L$ is added with a large positive mass-squared.  The mechanism 
of Ref.~\cite{m01} is then applicable. Let the trilinear coupling 
connecting the SM Higgs doublet $\Phi=(\phi^+,\phi^0)$ to $\eta$ and 
$\chi^0$ be $\mu \chi^0 \Phi \eta$, then the induced vacuum expectation 
value (VEV) of the heavy $\eta^0$ is given by 
$-\mu \langle \chi^0 \rangle \langle \phi^0 \rangle/m^2_\eta$, which 
is very much suppressed.  Being proportional to $\langle \eta^0 \rangle$, 
the Dirac neutrino masses are thus naturally very small.  This mechanism 
has recently been invoked in a number of 
scenarios~\cite{m21-1,m21-2,m21-3,m21-4}. 

The $U(1)_L$ gauge symmetry is broken to $Z^L_3$, with each particle 
transforming as $\omega^L$ with $\omega^3=1$.   
As a bonus, some of the new fermions added to render $U(1)_L$ anomaly-free 
form a dark sector~\cite{bh18}.  The dark symmetry is derived~\cite{m15,m20} 
from lepton symmetry by multiplying with $\omega^{-2j}$, where $j$ is the 
particle's intrinsic spin.

The new particles, including the $U(1)_L$ gauge boson $Z_L$, are not easily 
produced.  They may be rather light, of order 100 GeV, and have escaped 
detection.  In particular, the dark sector may explain the discrepancy 
of the experimental measurement of the muon anomalous magnetic moment 
compared against the theoretical SM prediction.

\noindent \underline{\it Model}~:~ 
The relevant particles of this model are listed in Table~1. The SM leptons 
come in 3 families; the other fermions and scalars are not duplicated. The 
$U(1)_L$ charge of $\nu_R$ is chosen to allow it to have a naturally small 
mass because it couples to $\eta$ not $\Phi$.
\begin{table}[tbh]
\centering
\begin{tabular}{|c|c|c|c|c|c|}
\hline
fermion/scalar & $SU(2)$ & $U(1)_Y$ & $U(1)_L$ & $Z_3^L$ & $Z_3^D$ \\
\hline
$(\nu,e)_L$ & $2$ & $-1/2$ & $1$ & $\omega$ & $1$ \\ 
$e_R$ & $1$ & $-1$ & $1$ & $\omega$ & $1$ \\ 
$\nu_R$ & $1$ & $0$ & $-2$ & $\omega$ & $1$ \\ 
\hline
$(N^0,E^-)_L$ & $2$ & $-1/2$ & $-3$ & $1$ & $\omega^{-1}$ \\ 
$(N^0,E^-)_R$ & $2$ & $-1/2$ & $0$ & $1$ & $\omega^{-1}$ \\
\hline
$F^-_R$ & $1$ & $-1$ & $-3$ & $1$ & $\omega^{-1}$ \\
$F^-_L$ & $1$ & $-1$ & $0$ & $1$ & $\omega^{-1}$ \\
\hline
$n_{1L}$ & $1$ & $0$ & $-5$ & $\omega$ & $1$ \\
$n_{1R}$ & $1$ & $0$ & $4$ & $\omega$ & $1$ \\
\hline
$n_{2L}$ & $1$ & $0$ & $-3$ & $1$ & $\omega^{-1}$ \\
$n_{2R}$ & $1$ & $0$ & $-6$ & $1$ & $\omega^{-1}$ \\
\hline
\hline
$(\phi^+,\phi^0)$ & $2$ & $1/2$ & $0$ & $1$ & $1$  \\ 
$(\eta^+,\eta^0)$ & $2$ & $1/2$ & $-3$ & $1$ & $1$ \\
$\chi_1^0$ & $1$ & $0$ & $3$ & $1$ & $1$ \\
$\chi_2^0$ & $1$ & $0$ & $-9$ & $1$ & $1$ \\
\hline
$\zeta^0$ & $1$ & $0$ & $1$ & $\omega$ & $\omega$ \\
\hline
\end{tabular}
\caption{Fermions and scalars under gauge $U(1)_L$ with resulting $Z_3^L$  
and $Z_3^D$.}
\end{table}

This model is free of anomalies: 
\begin{eqnarray}
[SU(2)]^2 U(1)_L &:& 3 [(1/2)(1)] + (1/2)[-3-0] = 0, \\~ 
[U(1)_Y]^2 U(1)_L &:& 3 [2(-1/2)^2 - (-1)^2] + 2(-1/2)^2 [-3-0] 
+(-1)^2 [0-(-3)] = 0, \\ 
U(1)_Y [U(1)_L]^2 &:& 3 [2(-1/2) - (-1)] + 2(-1/2)[(-3)^2 - 0] 
-[0 -(-3)^2] = 0, \\~ 
[U(1)_L]^3 &:& 3 [2-1-(-2)^3] + 2[(-3)^3 - 0] + [0 - (-3)^3] \nonumber \\ 
&& + [(-5)^3 - (4)^3] + [(-3)^3 - (-6)^3] = 0, \\ 
U(1)_L &:& 3 [2 - 1 - (-2)] + 2 [-3-0] + [0-(-3)] \nonumber \\ 
&& + [-5-4] + [-3-(-6)] = 0.
\end{eqnarray}

The allowed Yukawa terms are:
\begin{eqnarray}
&& (\bar{\nu}_L \phi^+ + \bar{e}_L \phi^0) e_R, ~~~ \bar{\nu}_R 
(\nu_L \eta^0 - e_L \eta^+), ~~~ \bar{\nu}_R n_{1L} \chi_1, ~~~ 
\bar{n}_{1L} n_{1R} \chi_2, \\
&& \bar{F}_L F_R \chi_1, ~~~ (\bar{N}_R N_L + \bar{E}_R E_L) \chi_1, ~~~ 
(\bar{N}_R \phi^+ + \bar{E}_R \phi^0) F_L, ~~~ 
(\bar{N}_L \phi^+ + \bar{E}_L \phi^0) F_R, \\ 
&& \bar{n}_{2L} n_{2R} \chi_1, ~~~ \bar{n}_{2L} (N_R \eta^0 - E_R \eta^+), ~~~ 
\bar{n}_{2R} (N_L \eta^0 - E_L \eta^+), \\ 
&& \bar{e}_R F_L \zeta, ~~~ (\bar{\nu}_L N_R + \bar{e}_L E_R) \zeta, ~~~ 
\bar{\nu}_R n_{2L} \zeta. 
\end{eqnarray}
After symmetry breaking by $\langle \chi_1 \rangle = u_1$ and 
$\langle \chi_2 \rangle = u_2$, the terms involving the SM leptons show 
that the Dirac fermion $n_1$ mixes with one linear combination 
of $\nu_R$.  This means that the three $\nu_L$ of the SM pair up with 
three linear combinations out of the four right-handed singlets, i.e. 
the three $\nu_R$ plus $n_{1R}$.  The remaining combination pairs up with 
$n_{1L}$.

The Dirac fermions $(N,E)$, $F$, and $n_2$ mix among themselves, but are 
distinct from the SM leptons and $n_1$.  The two sectors are connected 
only through $\zeta$.  The residual symmetry is $Z^L_3$ as shown in Table~1. 
The dark symmetry $Z^D_3$ is obtained by multiplying each with 
$\omega^{-2j}$,  where $\omega^3=1$ and $j$ is the particle's intrinsic 
spin.  The lightest particle of the dark sector is assumed to be $\zeta$.

\noindent \underline{\it Muon Anomalous Magnetic Moment}~:~
The new interactions which contribute to the muon anomalous magnetic moment 
are
\begin{equation}
{\cal L}_{int} = g_L Z_L^\alpha \bar{\mu} \gamma_\alpha \mu + [y_E^\mu \zeta 
\bar{\mu}_L E_R + y_F^\mu \zeta \bar{\mu}_R F_L + H.c.],
\end{equation}
with the results~\cite{qs14}:
\begin{equation}
\Delta a_\mu (Z_L) = {g_L^2 m_\mu^2 \over 12 \pi^2 m^2_{Z_L}} = 
{m_\mu^2 \over 216 \pi^2 (u_1^2 + 9 u_2^2)}, 
\end{equation}
\begin{equation} 
\Delta a_\mu (\zeta,E) = {(y_E^\mu)^2 m_\mu^2 f(m_E^2/m_\zeta^2) \over 
96 \pi^2 m_\zeta^2}, ~~~
\Delta a_\mu (\zeta,F) = {(y_F^\mu)^2 m_\mu^2 f(m_F^2/m_\zeta^2) \over 
96 \pi^2 m_\zeta^2},
\end{equation}
where $m^2_{Z_L} = 18 g_L^2 u_1^2 + 162 g_L^2 u_2^2$ and 
\begin{equation}
f(r) = (r-1)^{-4} (r^3 - 6r^2 + 3r + 2 + 6r \ln r).
\end{equation}
As a numerical example, let $m_\zeta = 70$ GeV, $m_E = m_F = 110$ GeV, 
$y_E^\mu = y_F^\mu = 1.2$, and $u_1 = 92$ GeV with $u_2$ negligibly smaller, 
then the sum of the above yields $2.51 \times 10^{-9}$, which is the 
central value of the deviation of the world average experimental 
value~\cite{mug21} from the theoretical prediction~\cite{mug20} of the SM.  
The $m_E,m_F,m_\zeta$ values are chosen to avoid constraints from 
collider data~\cite{atlas14} on colorless charged 
particles decaying to muons and missing energy.  The $Z_L$ mass is  
$g_L$ times 390 GeV.  Since $Z_L$ couples only to leptons, the bound on 
$m_{Z_L}$ is 209 GeV, coming from the highest energy of the LEP II 
$e^+e^-$ collider.  This means that $g_L > 0.54$.  Also, $m_E = y_E u_1$ 
and $m_F = y_F u_1$, so $y_E = y_F = 1.2$ in this example.

\noindent \underline{\it Scalar Sector}~:~
The scalar potential consisting of $\Phi,\eta,\chi_{1,2},\zeta$ is given by
\begin{eqnarray}
V &=& -\mu_0^2 \Phi^\dagger \Phi + m_1^2 \eta^\dagger \eta - \mu_1^2 \chi_1^* 
\chi + m_2^2 \chi_2^* \chi_2 + m_3^2 \zeta^* \zeta \nonumber \\ 
&+& [\mu' \chi_1 \Phi^\dagger \eta + f \chi_2 \chi_1^3 + f' \chi_1^* \zeta^3 
+ H.c.] \nonumber \\ 
&+& {\rm the~usual~quartic~terms.}
\end{eqnarray}
The $\mu'$ and $f$ terms allow $\langle \eta^0 \rangle = v_1$ and 
$\langle \chi_2 \rangle = u_2$ to be suppressed from $m_1^2 > 0$ and 
$m_2^2 > 0$, using the mechanism of Ref.~\cite{m01}, i.e.
\begin{equation}
v_1 \simeq {-\mu' u_1 v_0 \over m_1^2}, ~~~ u_2 \simeq {-f u_1^3 \over m_2^2},
\end{equation}
where $\langle \phi^0 \rangle = v_0$ and $\langle \chi_1 \rangle = u_1$. 
The $f'$ term allows $\zeta$ to transform as $\omega$ under $Z_3$ after 
$\chi_1$ picks up a VEV.

The physical scalars after the spontaneous breaking of 
$SU(2)_L \times U(1)_Y \times U(1)_L$ are then the SM Higgs boson $h$, the 
corresponding $H$ from $U(1)_L$, and $\zeta$, whereas the  
$\eta$ doublet and $\chi_2$ singlet are much heavier.  Since $\zeta$ is 
assumed to be dark matter, its interactions with $h$ and $H$ are important 
considerations, i.e.
\begin{equation}
-{\cal L}_{int} = \lambda (\sqrt{2} v_0 h + {1 \over 2} h^2) \zeta^* \zeta + 
\lambda' (\sqrt{2} u_1 H + {1 \over 2} H^2) \zeta^* \zeta.
\end{equation}
The constraints on $\lambda$ and $\lambda'$ will be discussed later.

\noindent \underline{\it Neutrino Sector}~:~
There are 4 Dirac neutrinos in this model.  The $n_{1L}$ singlet pairs up 
with one linear combination of the three right-handed neutrinos (call it 
$\nu_R$) with mass proportional to $u_1$.  This $\nu_R$ will connect to 
one linear combination of the three left-handed neutrinos (call it $\nu_L$). 
The $2 \times 2$ mass matrix linking $(\nu_L, n_{1L})$ to $(n_{1R}, \nu_R)$ 
is of the form
\begin{equation}
{\cal M}_{\nu,n} = \pmatrix{0 & f_1 v_1 \cr f_2 u_2 & f_3 u_1}.
\end{equation}
Since $v_1$ and $u_2$ are suppressed, $\nu_L$ actually pairs up with $n_{1R}$ 
to form a very light Dirac neutrino of mass $(f_1 v_1)(f_2 u_2)/(f_3 u_1)$. 
This is small compared to the other two Dirac neutrino masses which are 
proportional to $v_1$ alone.  Hence this model predicts one nearly massless 
Dirac neutrino, two of order 1 eV, and one of order 100 GeV, which decays 
through $Z_L$ to SM leptons through $\nu_R-n_{1R}$ mixing.

\noindent \underline{\it Dark Sector}~:~
The dark fermions $N,E,F$ all decay to $\zeta^*$ and a lepton.  They are 
produced only through electroweak interactions.  As noted earlier, for 
$m_E=m_F=110$ GeV, their decay to muons with the large missing energy of 
$m_\zeta=70$ GeV is allowed by the present collider data~\cite{atlas14}. 
The interactions of $\zeta$ with the SM particles are shown in 
Eqs.~(10),(17), and through the $Z_L$ gauge boson.  In underground 
direct-search experiments using nuclear recoil, only the Higgs exchange 
is applicable.

The spin-independent cross section for elastic scattering off a xenon 
nucleus is
\begin{equation}
\sigma_0 = {1 \over \pi} \left( {m_\zeta m_{Xe} \over m_\zeta + m_{Xe}} 
\right)^2 \left| {54 f_p + 77 f_n \over 131} \right|^2,
\end{equation}
where~\cite{hint11}
\begin{eqnarray}
{f_p \over m_p} &=& \left[0.075 + {2 \over 27}(1-0.075) \right] 
{\lambda \over m_h^2 m_\zeta}, \\
{f_n \over m_n} &=& \left[0.078 + {2 \over 27}(1-0.078) \right]
{\lambda \over m_h^2 m_\zeta}.
\end{eqnarray}
For $m_\zeta = 70$ GeV, $m_h = 125$ GeV, and $m_{Xe}=122.3$ GeV, the upper 
limit on $\sigma_0$ is~\cite{xenon18} $7 \times 10^{-47}~{\rm cm}^2$, 
thereby yielding $\lambda < 1.36 \times 10^{-4}$.  

As for relic abundance, the cross sections of $\zeta \zeta^*$ to lepton 
pairs through $Z_L$ or Eq.~(10) are helicity-suppressed.  Assuming $H$ to 
be lighter than $\zeta$, the $\zeta \zeta^* \to HH$ amplitudes are shown 
in Fig.~1.
\begin{figure}[htb]
\vspace*{-5cm}
\hspace*{-3cm}
\includegraphics[scale=1.0]{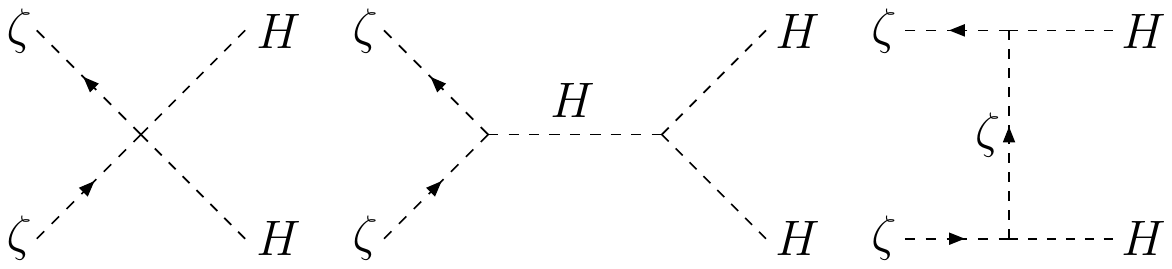}
\vspace*{-21.5cm}
\caption{Annihilation of $\zeta \zeta^* \to HH$.}
\end{figure}
The sum of these diagrams (with a factor of 2 in the last one) for 
$\zeta \zeta^*$ annihilation at rest is
\begin{equation}
\lambda_{eff} = \lambda' \left[ 1 + {3m_H^2 \over 4m^2_\zeta - m_H^2} - 
{4\lambda' u_1^2 \over 2m^2_\zeta - m_H^2} \right] 
= \lambda' (1.675 - 5.461 \lambda'),
\end{equation}
where $m_\zeta=70$ GeV, $m_H=60$ GeV, and $u_1=92$ GeV have been used. 
The annihilation cross section times relative velocity is
\begin{equation}
\sigma_{ann} \times v_{rel} = {\lambda^2_{eff} \over 64 \pi} 
{\sqrt{m_\zeta^2 - m_H^2} \over m^3_\zeta}.
\end{equation}
Setting this equal to the typical value of 
$3 \times 10^{-26}~{\rm cm}^3/{\rm s}$ for the correct relic abundance, 
$\lambda_{eff} = 0.07$ is obtained, for which $\lambda' = 0.05$ is a solution.
Once produced, $H$ decays through its very small mixing with $h$ to SM 
particles.

\noindent \underline{\it Concluding Remarks}~:~
Three seemingly unrelated fundamental issues in particle and astroparticle 
physics are shown to be connected in a proposed theoretical framework 
of gauged lepton number $U(1)_L$.  The three topics are (1) naturally small 
Dirac neutrino masses, (2) dark matter, and (3) the muon anomalous magnetic 
moment.  Assigning charges to leptons under $U(1)_L$, together with the 
addition of new fermions to render the theory anomaly-free as shown in 
Table~1, it is discovered that the breaking of gauge $U(1)_L$ results in 
a conserved $Z_3^L$ symmetry.  A dark sector also automatically appears, 
with dark $Z_3^D$ symmetry obtained by multiplying with $\omega^{-2j}$ 
for each particle, where $\omega^3=1$ and $j$ is the particle's intrinsic 
spin.  

This breaking of $U(1)_L$ by 3 units is also the key to having 
naturally small Dirac neutrino masses using the mechanism of Ref.~\cite{m01}. 
New particles transforming under $U(1)_L$ are not easily produced.  They 
may be below 100 GeV and have escaped detection.  Hence they are also 
suitable for explaining the discrepancy of the measured muon anomalous 
magnetic moment compared against the SM theoretical prediction.  Here 
the contributions come mainly from the dark sector, with a smaller share 
from the $Z_L$ gauge boson.   To summarize, the numerical values used 
in this study are $m_\zeta=70$ GeV for the scalar dark-matter candidate, 
$m_H=60$ GeV for the Higgs boson associated with $U(1)_L$ which is broken by 
$\langle \chi_1 \rangle = u_1 = 92$ GeV, and $m_E=m_F=110$ GeV for the 
charged fermions in the dark sector.

\noindent \underline{\it Acknowledgement}~:~
This work was supported 
in part by the U.~S.~Department of Energy Grant No. DE-SC0008541.

\bibliographystyle{unsrt}

\end{document}